\documentclass[12pt,nofootinbib]{revtex4}
\usepackage{amssymb}
\usepackage{amsmath, graphicx}
\usepackage{slashed}
\usepackage[unicode]{hyperref}
\usepackage{csquotes}
\usepackage{braket}
\usepackage{natbib}
\usepackage{array}
\usepackage{xcolor}

\begin{document}
\title{Gravitational waves from burdened primordial black holes dark matter}
\author{Ngo Phuc Duc Loc}
\thanks{Email: locngo148@gmail.com}

\affiliation{Department of Physics and Astronomy, University of New Mexico, Albuquerque, NM 87131, USA}

\begin{abstract}
Primordial black holes (PBHs) are the natural candidate of dark matter (DM) as they only interact gravitationally and can evade any experiments on earth. In the standard semiclassical calculation of Hawking radiation, PBHs with mass below $10^{15}\rm g$ evaporated by now and therefore cannot be DM. However, the recently-discovered quantum memory burden effect can significantly suppress the evaporation of PBHs after the half-decay time. This quantum effect could open up a new mass window below $10^{10} \rm g$ where PBHs can still exist today and be DM. In this paper, we compute the gravitational wave (GW) signals associated with the formation of PBHs in this new mass window. We consider two formation scenarios: PBHs formed from inflationary perturbation and PBHs formed from collapse of Fermi-balls in a first-order phase transition (FOPT). GWs produced from these two scenarios have distinct features and, while the GW from inflation peaks at high frequency, the GW from FOPT peaks at lower frequency that can be within the reach of future experiments.
\end{abstract}
\maketitle

\section{Introduction}

Despite various astrophysical evidences of dark matter (DM) \cite{Bertone:2016nfn}, we have not seen any evidence of this elusive matter here on earth. What we know is that DM is cold (i.e. moves slowly) and seems to only have gravitational interaction, so it can evade all the experiments on earth which rely on the non-gravitational interaction between DM and the visible, normal matter.

Remarkably, there is a kind of objects in the Universe that behave exactly like DM: black holes. Black holes move slowly and only have gravitational interaction with everything else. Of course, if black holes are indeed DM, there is less hope to detect DM directly on earth. This justifies the lack of earth-based experimental evidence of DM.

But we also know that DM must have already existed in the very early Universe. It cannot be something that just popped out recently. So if we want to invoke black holes as a solution of DM, they cannot be the usual kind of astrophysical black holes formed from supernovae. Instead, they must have a primordial origin. We call them primordial black holes (PBHs).

If PBH is formed from collapse of overdense region produced by inflationary perturbation, its mass is set by the scale of the Hubble horizon at the formation time \cite{Carr:2020gox}. Consequently, the upper bound on the energy scale of inflation inferred from the cosmic microwave background (CMB) radiation, which is around $10^{13}\ \rm GeV$, implies a lower bound on the mass of PBH: $M_{\rm PBH}>1 \rm g$. According to the standard semiclassical calculation \cite{Hawking:1975vcx}, PBHs in the mass range $1 \rm g\lesssim M_{\rm PBH}\lesssim 10^9 \rm g$ evaporated before the Big Bang Nucleosynthesis (BBN) and therefore are not subject to any constraint on their population. PBHs in this mass window cannot be DM (because we do see DM today), but their evaporation product could be DM. PBHs in the mass range $10^9 \rm g\lesssim M_{\rm PBH}\lesssim 10^{17}\rm g$ evaporated between the BBN epoch and the present, so their population is constrained by multiple astrophysical data. PBHs in the mass range $10^{17} \rm g\lesssim M_{\rm PBH}\lesssim 10^{22} \rm g$ can be the entire DM; we call this the asteroid-mass window as a typical asteroid observed in the solar system is in this mass range. They are heavy enough to not yet evaporate and they are small enough to evade gravitational lensing experiment. PBHs heavier than $10^{22} \rm g$ cannot be the entire DM as their population is constrained by lensing, accretion, and dynamical effects. Of course, one can also imagine a hybrid solution to the DM problem where PBHs together with some particle DM candidate could comprise the total observed DM abundance.

However, it was recently shown that quantum memory burden effect could significantly suppress the evaporation of black hole after its half-decay time \cite{Dvali:2020wft,Dvali:2024hsb}. The half-decay time is defined as the time when the black hole has lost half of its initial mass due to evaporation and its Bekeinsgtein-Hawking entropy, which is defined as $S_{\rm BH}=4\pi GM^2$ in natural units \cite{Bekenstein:1973ur}, reduces to a quarter of its initial value: $M\rightarrow M/2,\ S_{\rm BH}\rightarrow S_{\rm BH}/4$. The basic idea is that, because the information capacity of the black hole is reduced, some initially-gapless memory modes will acquire energy gaps due to the reduction of the number of soft gravitons. As a result, it costs much more energy to emit any further Hawking radiation and therefore creates a ``burden" that prevents further decay of black hole. In particular, the evaporation rate is suppressed by a factor $S_{\rm BH}^{-k}$, where $k\geq 0$ and is a free parameter. As shown in \cite{Alexandre:2024nuo,Thoss:2024hsr}, this effect opens up a new mass window below $10^{10}\rm g$ where PBH evaporates partially until its half-decay time and then becomes stabilized to be DM.

If this novel quantum effect of black hole turns out to be true, it would be natural and interesting to study various resulting phenomenologies. For example, DM production or baryogenesis from evaporation of burdened PBHs has been studied in \cite{Haque:2024eyh,Barman:2024iht}. A PBH dominated era can also produce gravitational wave (GW) due to number density fluctuation when the PBHs evaporate \cite{Papanikolaou:2020qtd,Papanikolaou:2022chm}, which is useful in probing primordial non-Gaussianity \cite{Papanikolaou:2024kjb,He:2024luf}, and so modified Hawking evaporation would also lead to modified GW spectrum \cite{Balaji:2024hpu,Bhaumik:2024qzd}.

In this paper, we wish to compute the GW spectrum associated with the formation of PBHs in the new mass window below $10^{10}\rm g$ where PBHs can be the entire DM thanks to the quantum memory burden effect. There are various physical mechanisms to form PBHs in the early Universe. The most trivial way is through the collapse of sufficiently large inflationary perturbation \cite{Carr:1974nx}. Another natural way to form PBHs is through the collapse of Fermi-balls in a first-order phase transition (FOPT) \cite{Kawana:2021tde}. Both of these formation scenarios could produce the corresponding GW signal that may be observed in future experiments \cite{Ghosh:2022okj,Huang:2022him,Xie:2023cwi}\footnote{These two PBH formation scenarios are just a few among many other possible mechanisms such as PBHs formed from quark confinement \cite{Dvali:2021byy}, PBHs formation in the context of loop quantum cosmology \cite{Papanikolaou:2023crz} or F(R) gravity \cite{Banerjee:2022xft}, and many others. Nevertheless, the two formation scenarios from inflation and FOPT that we consider here are usually recognized as the most popular and well-motivated ones.}. The detection of such GW signal could be an evidence that supports the existence of PBHs in this new mass window and explains the DM problem\footnote{Other complimentary studies such as neutrinos or cosmic rays emission from burdened PBHs can also be useful to conclusively confirm the existence of such PBHs \cite{Chianese:2024rsn,Zantedeschi:2024ram}.}.

The rest of this paper is organized as follows. In Sec.\ref{sec:PBH formation}, we show the mass and the population of PBHs formed from two different mechanisms: inflationary perturbation and collapse of Fermi ball from a FOPT. The calculations will take into account a correction due to partially evaporated black holes. In Sec.\ref{sec:GWs}, we compute the GW spectrum corresponding to the new mass window below $10^{10}\rm g$ where PBHs can be the entire DM thanks to the memory burden effect. As we will see, GWs spectrum produced from inflation and FOPT have distinct features and, while the former peaks at the high-frequency regime, the latter peaks at the lower frequency regime that can be probed by upcoming experiments. We summarize our results in Sec.\ref{sec:conclusion}. The redshift business of GW spectrum in generic cosmological background is reviewed in Appendix A. Comparison between different contributions to the total GW spectrum from FOPT is given in Appendix B.

We note that, while we were preparing our paper, we were aware of \cite{Kohri:2024qpd,Barker:2024mpz,Jiang:2024aju} that also study scalar-induced GW from burdened PBH DM\footnote{See also \cite{Franciolini:2023osw} for GW associated with the formation of microscopic DM relics, including the Hawking remnants after evaporation.}, but they did not discuss about the GW from a FOPT scenario.

\section{Mass and population of primordial black holes}\label{sec:PBH formation}

In this section, we discuss the mass and population of PBHs in the two formation scenarios: PBHs formed from inflationary perturbation and PBHs formed from collapse of Fermi-balls in a FOPT. We assume PBHs form in the standard thermal history with the radiation-dominated (RD) phase before BBN and they have an almost monochromatic mass function. Unlike the standard results in the literature, we will now take into account the fact that PBH becomes stabilized after its half-decay time due to the memory burden effect.

\subsection{PBHs formed from inflationary perturbation}\label{subsec: PBH formed from inflation}
If PBH was formed from inflationary perturbation, its initial mass is typically set by the scale of the horizon mass \cite{Carr:2020gox}:
\begin{equation}
    M_{\rm PBH,i}\simeq \gamma M_{\rm H,i},
\end{equation}
where $\gamma\simeq 0.2$, $M_{\rm H,i}=\sqrt{45}m_{\rm pl}^3/4\pi^{3/2}g_{\rho ,\rm i}^{1/2}T_{\rm i}^2$ is the horizon mass at the formation time with $m_{\rm pl}\sim 10^{19}\ \rm GeV$ is the Planck mass and $g_{\rho ,\rm i}$ is the effective relativistic degrees of freedom (d.o.f.) at temperature $T_{\rm i}$. This can also be written in a normalized form as
\begin{equation}
    \left(\frac{M_{\rm PBH,i}}{\rm g}\right)\approx 1.89\times 10^{31}\left(\frac{\gamma}{0.2}\right)
\left(\frac{106.75}{g_{\rho,\rm i}}\right)^{1/2}\left(\frac{\rm GeV}{T_{\rm i}}\right)^2.
\end{equation}
After the PBH evaporates to half of its initial mass $M_{\rm PBH}=M_{\rm PBH,i}/2$, the evaporation is suppressed and the PBH becomes stabilized. In what follows, the notation $M_{\rm PBH}$ will always refer to the final mass of PBH.

PBHs with mass less than $10^{9} \rm g$ evaporate before BBN in the RD phase. Because the population of PBHs must be subdominant at early time in order to explain the correct DM abundance, their partial evaporation does not lead to significant entropy injection. Therefore, the population of PBHs at the half-decay time is
\begin{equation}\label{eq:beta_half_1}
    \beta_{\rm PBH, half}=\frac{\rho_{\rm PBH,half}}{\rho_{\rm rad,half}}=\frac{\beta_{\rm PBH, i}}{2}\left(\frac{a_{\rm half}}{a_{\rm i}}\right),
\end{equation}
where $\beta_{\rm PBH,i}$ is the initial PBH's population. After the half-decay time, we can reshift the PBH's abundance to today and get
\begin{align}
        \beta_{\rm PBH, half}&=\frac{4}{3}\frac{M_{\rm PBH}n_{\rm PBH,half}}{T_{\rm half}s_{\rm half}}\\
        &=\frac{4}{3}\left(\frac{g_{\rm s,half}}{g_{\rm s,i}}\right)^{1/3}\left(\frac{a_{\rm half}}{a_{\rm i}}\right)\frac{1}{T_{\rm i}}\frac{M_{\rm PBH}n_{\rm PBH}(t_0)}{s(t_0)}\\
        &\simeq 6\times 10^{-26}\left(\frac{g_{\rm s,half}}{g_{\rm s,i}}\right)^{1/3}\left(\frac{a_{\rm half}}{a_{\rm i}}\right)\left(\frac{g_{\rho ,\rm i}}{106.75}\right)^{1/4}\gamma^{-1/2}\left(\frac{M_{\rm PBH,i}}{\rm g}\right)^{1/2}f_{\rm PBH}.\label{eq:_beta_half_2}
\end{align}
We used the fact that $\rho_{\rm rad}=3sT/4$, $T\propto g_{\rm s}^{-1/3}a^{-1}$, $n_{\rm PBH,half}/s_{\rm half}= n_{\rm PBH}(t_0)/s(t_0)$ with $t_0$ is the present time, and $f_{\rm PBH}\equiv\Omega_{\rm PBH}/\Omega_{\rm DM}$ is the current fraction of PBH in DM. PBHs can be the entire DM if $f_{\rm PBH}=1$. For the numerical factors, we used $m_{\rm pl}=1.22\times 10^{19}\ \rm GeV=2.18\times 10^{-5}\rm g$, $s(T_0)=2\pi^2g_{\rm s,0}T_0^3/45$ with $g_{\rm s,0}=3.94$ and $T_0=2.3\times 10^{-13}\ \rm GeV$, $\Omega_{\rm DM}=0.27$, $H_0=67\ \rm km/s/Mpc=1.43\times 10^{-42}\ \rm GeV$ \cite{Planck:2018vyg}.  We also distinguished between $g_\rho$, which refers to the effective relativistic d.o.f. in energy density, and $g_{\rm s}$, which refers to the effective relativistic d.o.f. in entropy density.

By using Eqs. \ref{eq:beta_half_1} and \ref{eq:_beta_half_2}, and $M_{\rm PBH,i}=2M_{\rm PBH}$, we can express the initial population of PBHs in terms of the final fraction of PBH in DM and PBH's mass:
\begin{align}
    \beta_{\rm PBH, i}&\simeq 2\sqrt{2}\left(\frac{g_{\rm s,half}}{g_{\rm s,i}}\right)^{1/3}6\times 10^{-26}\left(\frac{g_{\rho,\rm i}}{106.75}\right)^{1/4}\gamma^{-1/2}\left(\frac{M_{\rm PBH}}{\rm g}\right)^{1/2}f_{\rm PBH}\label{eq:beta_f_M gammaPT}\\
    &\approx 3.8\times 10^{-25}f_{\rm PBH}\left(\frac{M_{\rm PBH}}{\rm g}\right)^{1/2}.\label{eq:beta_f_M}
\end{align}
We used the fact that, at the high temperature we consider here, $(g_{\rm s,half}/g_{\rm s,i})^{1/3}\sim O(1)$, $g_{\rho,\rm i}\sim 106.75$, and $\gamma\simeq 0.2$. We see that the quantum memory burden effect gives a correction factor $2\sqrt{2}(g_{\rm s,half}/g_{\rm s,i})^{1/3}$ to the standard result. This implies that, for a given final mass and population of PBHs, the initial population must be slightly greater than the usual value to account for the partial evaporation of PBHs.

On the theoretical side, PBH will form during the RD phase if the density contrast is greater than the threshold value $\delta_c\approx 0.42$. Assuming a Gaussian distribution, the initial population of PBHs is given by \cite{Carr:1975qj}
\begin{equation}\label{eq: beta_sigmaH}
    \beta_{\rm PBH,i}=\int_{\delta_c}^{O(1)}\frac{1}{\sqrt{2\pi\sigma_H^2}}\exp\left(\frac{-\delta^2}{2\sigma_H^2}\right)d\delta\approx\rm  Erfc\left[\frac{\delta_c}{\sqrt{2}\sigma_H}\right],
\end{equation}
where $\rm Erfc[...]$ is the complementary error function and \cite{Kohri:2018qtx}
\begin{equation}
    \sigma_H(k)\simeq \frac{4}{9}\sqrt{P_\xi(k)}
\end{equation}
for a RD phase.

In order to compute the GW spectrum from scalar-induced perturbation, we need an explicit form of the power spectrum of curvature perturbation $P_\xi(k)$. If we simply extrapolate this spectrum from the CMB scale to smaller scale, the spectrum is slightly red-tilted at small scale and therefore cannot produce PBHs. In order to have a decent chance of creating a sufficient population of PBHs, we must assume a blue-tilted power spectrum at small scale. There are many inflation models that could do this and we have no wish to invoke any specific model. Instead, we will choose two simple, phenomenological forms of the power spectrum.

We can parameterize the power spectrum of curvature perturbation as follows:
\begin{equation}\label{eq:power law spectrum}
P_{\xi}(k)=
    \begin{cases}
        A_{\rm s}\left(\frac{k}{k_*}\right)^{n_{\rm s}-1+\frac{\alpha_{\rm s}}{2}\ln\left(\frac{k}{k_*}\right)}\hspace{2cm}k\leq k_{\rm p}\\
        P_{\xi,\rm max}\left(\frac{k_{\rm p}}{k}\right)^2\hspace{3.5cm} k\geq k_{\rm p}
    \end{cases},
\end{equation}
where $k_*$ is the pivot scale where we have data from the CMB, $A_{\rm s}$ is the magnitude of scalar perturbation, $n_{\rm s}$ is the tilt of the spectrum, $\alpha_{\rm s}$ is the running spectral index, and $k_{\rm p}$ is the peak mode where PBHs form. We assume the running of the running is zero for simplicity. At the pivot scale $k_*=0.05\ \rm Mpc^{-1}$, we have $A_{\rm s}=10^{-10}\exp (3.045)$ and $n_s=0.9625$ at $68\%$ C.L. \cite{Planck:2018jri}. The allowed parameter space for the running spectral index is $\alpha_{\rm s}=0.011\pm 0.021$ at $95\%$ C.L. \cite{Cabass:2016ldu}. In principle, one can consider an abrupt cutoff of the spectrum at the peak scale $k_{\rm p}$ where PBHs form. However, a smooth cutoff is more natural and we chose the form $\propto k^{-2}$ for $k\geq k_{\rm p}$ following \cite{Kohri:2018qtx}. This narrowly peaked power spectrum, together with the fact that the PBH population is exponentially sensitive to the spectrum, induces an almost monochromatic mass function at $k_p$.

Another realistic form of the power spectrum is the log-normal form:
\begin{equation}\label{eq:log normal spectrum}
    P_\xi(k)=\frac{A}{\sqrt{2\pi\sigma^2}}\exp\left[-\frac{\ln^2(k/k_p)}{2\sigma^2}\right],
\end{equation}
where $A$ sets the magnitude and $\sigma$ sets the width of the power spectrum. This form features a symmetric peak around the time of PBH formation. For single-field inflation, it is known that the steepest increase of the curvature power spectrum is $\propto k^{4}$ corresponding to $\sigma\gtrsim 1$ \cite{Byrnes:2018txb}. Additionally, it is also known that, if $\sigma\lesssim 2$, the mass function induced from this spectrum will be almost monochromatic and one can therefore compare the result with the constraints on monochromatic PBH mass function \cite{Kozaczuk:2021wcl}. Thus, we will choose $\sigma=1$ in this paper.

The peak wavenumber corresponding to the formation of PBHs can be calculated as
\begin{align}
        k_p&=a_iH_i\\
        &=\left(\frac{g_{s,0}}{g_{s,i}}\right)^{1/3}\frac{T_0}{T_i}H_i\\
        &=\left(\frac{g_{s,0}}{g_{s,i}}\right)^{1/3}\frac{T_0\pi^{3/4}g_{\rho,i}^{1/4}\gamma^{1/2}}{45^{1/4}}\left(\frac{m_{\rm pl}}{M_{\rm PBH,i}}\right)^{1/2}\\
        &\approx 7.3 \times 10^{22}\ \rm Mpc^{-1}\left(\frac{\rm g}{M_{\rm PBH,i}}\right)^{1/2}\\
        &\approx 5.16\times 10^{22}\ \rm Mpc^{-1}\left(\frac{\rm g}{M_{\rm PBH}}\right)^{1/2}.\label{eq:kp mass}
\end{align}
We can also translate the wavenumber into the frequency language as
\begin{equation}\label{eq:frequency wave number}
    f=\frac{c}{2\pi}k\approx 1.55\times 10^{-15}\ \rm Hz\left(\frac{k}{\rm Mpc^{-1}}\right).
\end{equation}

By using Eqs. \ref{eq:beta_f_M}, \ref{eq: beta_sigmaH}, \ref{eq:kp mass}, together with the power spectrum forms in Eq. \ref{eq:power law spectrum} or Eq. \ref{eq:log normal spectrum},  we can find the corresponding parameters of the spectrum that produces $f_{\rm PBH}=1$ for a given PBH's mass. The results are shown in Table \ref{tab:data}. We use these parameters for our computation of GW spectrum later on.

\begin{table}[h!]
    \centering
\begin{tabular}{|>{\centering} m{3cm}|>{\centering} m{3cm}|>{\centering} m{3cm}|>{\centering} m{3cm}| m{3cm}<{\centering}|}
\hline
     $M_{\rm PBH}(\rm g)$ & $P_{\xi,\rm max}$ & $k_p(\rm Mpc^{-1})$ & $\alpha_s$ & $A(\sigma =1)$  \\
    \hline
    $10^3$ & 0.00889 & $1.63\times 10^{21}$ & 0.0128 & 0.02228\\
    \hline
    $10^6$ & 0.00954 & $5.16\times 10^{19}$ & 0.0146 & 0.02391\\
    \hline
    $10^9$ & 0.01029 & $1.63\times 10^{18}$ & 0.0169 & 0.02579\\
    \hline
\end{tabular}
    \caption{Parameters of curvature power spectrum that can produce $f_{\rm PBH}=1$ for different PBH's masses.}
    \label{tab:data}
\end{table}

\subsection{PBHs formed from the collapse of Fermi balls during a FOPT}

In a FOPT, the bubbles of the true vacuum nucleate and particles inside the bubbles acquire mass. However, if the massless fermions, $\chi$, in the false vacuum have much less thermal kinetic energy than their mass, they cannot penetrate into the bubble and get trapped in the false vacuum. The bubbles will expand and eventually fill the entire Universe. If there is an asymmetry between the trapped fermions and antifermions such that there is a net number of $\chi$ that survive, the Fermi balls will form. Inside the Fermi balls, $\chi$ interact with each other via attractive Yukawa potential. As the Universe cools down, the range of force increases and, when it becomes comparable to the mean separation of the fermions, the Fermi balls will collapse to form PBHs \cite{Kawana:2021tde}.

If PBHs were formed from the collapse of Fermi-balls in a FOPT, their initial mass is given by \cite{Kawana:2021tde}
\begin{equation}\label{eq:M_FOPT}
    M_{\rm PBH,i}\approx 1.4\times 10^{21}{\rm g}\  v_w^3\left(\frac{\eta_\chi}{10^{-3}}\right)\left(\frac{100}{g_{*}}\right)^{1/4}\left(\frac{100\ \rm GeV}{T_{*}}\right)^2\left(\frac{100}{\beta/H_*}\right)^3\alpha_*^{1/4},
\end{equation}
and their initial population is given by
\begin{equation}\label{eq:beta_FOPT}
    \beta_{\rm PBH, i}'\approx 1.4\times 10^{-15}v_w^{-3}\left(\frac{g_{*}}{100}\right)^{1/2}\left(\frac{T_*}{100\ \rm GeV}\right)^3\left(\frac{\beta/H_*}{100}\right)^3\left(\frac{M_{\rm PBH,i}}{10^{15}\rm g}\right)^{3/2}.
\end{equation}
Here, $v_w$ is the velocity of bubble wall, $\eta_\chi$ is the asymmetry parameter of fermion defined as $\eta_\chi\equiv(n_\chi-n_{\Bar{\chi}})/s$, $T_*$ is the phase transition temperature with the associated effective d.o.f. $g_*$ at that time, $\beta/H_*$ is the relative time scale of phase transition to the Hubble expansion, and $\alpha_*$ is the ratio of vacuum energy density released to the radiation energy density of the thermal bath.

Notice that, for the FOPT case, it is more convenient to use the normalized initial population defined as \cite{Carr:2020gox}
\begin{equation}
    \beta_{\rm PBH,i}'\equiv \gamma_{\rm PT}^{1/2}\left(\frac{g_{\rho,\rm i}}{106.75}\right)^{-1/4}\left(\frac{h}{0.67}\right)^{-2}\beta_{\rm PBH,i}.
\end{equation}
Note that $\gamma_{\rm PT}$ is now not a constant but depends on phase transition's parameters like in Eq. \ref{eq:M_FOPT}. $h$ is the normalized Hubble constant and we can set it to be $h=0.67$ \cite{Planck:2018vyg}. By using Eq. \ref{eq:beta_f_M gammaPT}, we can again express the initial population of PBH in terms of its final abundance and mass as\footnote{Note that Eq. \ref{eq:beta_f_M gammaPT} is completely general as all we said is that $M_{\rm PBH}=\gamma M_{\rm H}$. In light of Eq. \ref{eq:M_FOPT}, we can replace $\gamma\rightarrow\gamma_{\rm PT}$ with $\gamma_{\rm PT}$ being whatever in front of $M_{\rm H}$.}:
\begin{align}
    \beta_{\rm PBH,i}'&\simeq 2\sqrt{2}\left(\frac{g_{\rm s,half}}{g_{\rm s,i}}\right)^{1/3}6\times 10^{-26}f_{\rm PBH}\left(\frac{M_{\rm PBH}}{\rm g}\right)^{1/2}\\
    &\approx 1.7\times 10^{-25}f_{\rm PBH}\left(\frac{M_{\rm PBH}}{\rm g}\right)^{1/2}.\label{eq:beta PBH prime}
\end{align}

\section{Gravitational waves}\label{sec:GWs}
In this section, we compute the GW spectrum associated with the formation of PBHs from either inflationary perturbation or collapse of Fermi-balls in a FOPT. We will focus on the the new mass window below $10^{10}\rm g$ where PBHs can be the entire DM thanks to the quantum memory burden effect. As mentioned in the introduction, the suppression of PBH's evaporation depends on the power $k$ of the Bekeinstein-Hawking entropy. The larger the power $k$, the stronger the suppression of Hawking evaporation. If $k=2$, PBHs in the mass range $10^5\rm g\lesssim M_{\rm PBH}\lesssim 10^{10}\rm g$ can be the entire DM. Whereas if $k=3$, the lower bound can be extended down to $\sim 10^3\rm g$, and so on. In order to keep the discussion general, we will not choose a specific value of $k$ but instead choose three benchmark points as an example: $M_{\rm PBH}=\{10^3\rm g,10^6\rm g,10^9\rm g\}$.

\subsection{Scalar-induced GW from inflation}

At first order in cosmological perturbation theory, the scalar, vector, and tensor modes are decoupled. They evolve independently. The vector mode decays as $\propto a^{-2}$ with $a$ being the scale factor, so we usually do not care about it. The magnitude of GW from the tensor mode depends on the tensor-to-scalar ratio, $r$. If $r$ is large, GW is large. The observational upper bound on $r$ from the CMB therefore puts a constraint on the GW produced from tensor modes \cite{Maggiore:2018sht}.

At second order in perturbation theory, however, the scalar and tensor modes are coupled to each other. Because we know that there is scalar perturbation imprinted on the CMB, we know that there must be also the associated GW, no matter how small they are. We call this scalar-induced GW (see \cite{Domenech:2021ztg,Yuan:2021qgz} for a review). As discussed in Sec.\ref{subsec: PBH formed from inflation}, we must have an enhanced scalar perturbation at small scale in order to produce PBHs. This enhanced scalar perturbation will, in turn, produce the associated GW that we are about to compute\footnote{It is insightful to note that, in contrast to the usual naive expectation (possibly coming from the intuition with linear physical theory), the GW produced at second order does not need to be smaller than GW produced at first order. This special feature comes from the nonlinear nature of general relativity which couples the scalar and tensor modes at higher order in perturbation theory. An enhanced scalar perturbation does not enhance GW at first order, but it can enhance GW at second order or higher.}.

The fractional energy density of GW at the emission time is \cite{Kohri:2018awv}
\begin{equation}
    \Omega_{\rm GW,i}(\eta,k)=\frac{1}{24}\left(\frac{k}{a(\eta)H(\eta)}\right)^2\overline{P_h(\eta,k)},
\end{equation}
where 
\begin{equation}
    P_h(\eta,k)=2\int_0^\infty dt\int_{-1}^1ds\left[\frac{t(2+t)(s^2-1)}{(1-s+t)(1+s+t)}\right]^2I^2_{\rm RD}(v,u,x)P_\xi (kv)P_\xi (ku),
\end{equation}
\begin{equation}
\begin{aligned}
    \overline{I^2_{\rm RD}(v,u,x\rightarrow\infty)}&=\frac{1}{2}\left[\frac{3(u^2+v^2-3)}{4u^3v^3x}\right]^2\Bigg[\left(-4uv+(u^2+v^2-3)\ln \Bigg|\frac{3-(u+v)^2}{3-(u-v)^2}\Bigg|\right)^2\\
    &+\pi^2(u^2+v^2-3)^2\Theta(v+u-\sqrt{3})\Bigg],
\end{aligned}
\end{equation}
\begin{equation}
    u\equiv\frac{t+s+1}{2};\hspace{1cm} v\equiv\frac{t-s+1}{2},
\end{equation}
and $x\equiv k\eta=k/a(\eta)H(\eta)$. The fractional energy density of GW today is (see Appendix A)
\begin{equation}\label{eq: Omega GW RD}
    \Omega_{\rm GW,0}\approx 1.58\times 10^{-4}g_{\rm \rho,i}^{-1/3}\Omega_{\rm GW,i}.
\end{equation}
We can use Eq. \ref{eq:frequency wave number} to translate our results into the frequency language.

\begin{figure}[h!]
    \centering
    \includegraphics[scale=0.85]{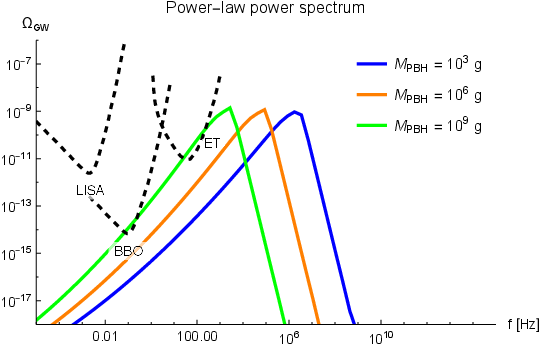}
    \includegraphics[scale=0.85]{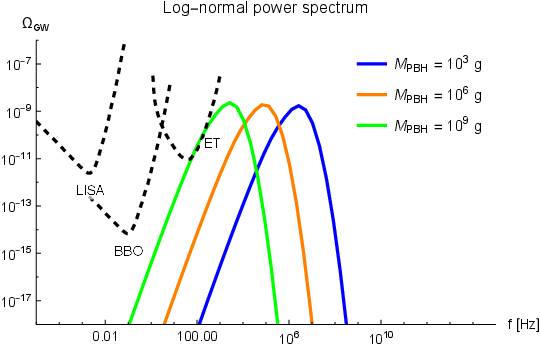}
    \caption{GW spectrum from inflationary perturbation for different PBH's masses. We choose $f_{\rm PBH}=1$ so that PBHs can be the entire DM. The left panel is for the power-law spectrum and the right panel is for the log-normal spectrum with $\sigma=1$. The black dashed contours represent sensitivity regions from different experiments.}
    \label{fig:GW inflation}
\end{figure}

Using Eqs. \ref{eq:power law spectrum} and \ref{eq:log normal spectrum}, we compute the GW signal as a function of frequency for different PBH's masses. We choose $f_{\rm PBH}=1$ so that PBHs can be the entire DM. In the high-frequency regime, there are some upcoming experiments that could be relevant for us such as LISA \cite{LISA:2017pwj}, BBO \cite{Yagi:2011wg}, DECIGO \cite{Kawamura:2011zz}, CE \cite{LIGOScientific:2016wof}, ET \cite{Punturo:2010zz}. The frequency range and sensitivity of DECIGO/CE are similar to that of BBO/ET respectively, so we do not show them here to avoid crowded figures. We show our results together with the sensitivity curves of some notable GW experiments\footnote{The GW sensitivity curves of LISA and BBO were taken from \cite{Moore:2014lga}, and that of ET was taken from \cite{ET:2019dnz}.} in Fig. \ref{fig:GW inflation}.


From Fig. \ref{fig:GW inflation}, we see that the GW spectrum peaks at frequency $\sim 10^3\ \rm Hz$ or above for the new mass window below $10^{10}\rm g$ where PBHs can be the entire DM opened up due to memory burden effect. The peak frequency $k_p$ is inversely proportional to the mass, which can also be seen in Eq. \ref{eq:kp mass}. The magnitude of GW signal at the peak is around $\sim 10^{-9}$. Recall that $\Omega_{\rm GW}$ must be less than $10^{-6}$ due to the BBN constraint on the effective relativistic d.o.f. Although the strength of our GW signal is reachable in future experiments, its peak frequency is out of reach of those experiments. Only a portion of the infrared tail of the GW spectrum may be observed \cite{Kohri:2024qpd}. Therefore, we need novel experiments that could probe very high-frequency GW.

\subsection{GW from FOPT}

In a FOPT, the energy gained in the transition from the false to the true vacuum is transferred into the kinetic energy of the bubble walls. When the bubbles expand and eventually collide, they release a huge amount of energy and some of them is in the form of GW. At the same time, the bubbles also interact with the cosmic fluid and, as they collide, they also produce turbulence and acoustic waves. In summary, GW from a FOPT can come from three different sources: bubble collision, sound wave, and turbulence \cite{Maggiore:2018sht}. Typically, the contribution from sound wave is most dominant \cite{Hindmarsh:2015qta}. The total fractional energy density of GW today is
\begin{equation}
    \Omega_{\rm GW}^{\rm PT}=\Omega_{\rm col}+\Omega_{\rm sw}+\Omega_{\rm turb},
\end{equation}
where $\Omega_{\rm col},\Omega_{\rm sw},\Omega_{\rm turb}$ are contributions from bubble collision, sound wave, and turbulence respectively.

In the following, we mostly follow \cite{Caprini:2015zlo} to have the form of GW spectrum from three different sources. By using Eqs. \ref{eq:M_FOPT} and \ref{eq:beta_FOPT}, we can conveniently express these contributions in terms of the PBH's mass and population. This will provide a more direct correlation between the GW signal and the corresponding PBH's features. It will also help us choose the best benchmark parameters for experimental prospects as well as make clearer comparison with the GW signal from inflation.

The GW spectrum from bubble collision is given by
\begin{align}
    \Omega_{\rm col}&\approx 3.72\times 10^{-5}\left(\frac{100}{g_*}\right)^{1/3}\left(\frac{\beta}{H_*}\right)^{-2}\left(\frac{\kappa_{\rm col}\alpha_*}{1+\alpha_*}\right)^2\frac{0.11v_w^3}{0.42+v_w^2}S_{\rm col}(f)\\
    &\approx 1.47\times 10^{-8} \alpha_*^{-1/2}\eta_\chi^{-2}\frac{v_w}{0.42+v_w^2}\left(\frac{\kappa_{\rm col}\alpha_*}{1+\alpha_*}\right)^2\left(\frac{g_*}{100}\right)^{-1/2}(\beta_{\rm PBH,i}')^{4/3}S_{\rm col}(f),
\end{align}
where the spectral shape is
\begin{equation}
    S_{\rm col}(f)=\frac{3.8(f/f_{\rm peak,col})^{2.8}}{1+2.8(f/f_{\rm peak,col})^{3.8}},
\end{equation}
the peak frequency is
\begin{align}
    f_{\rm peak,\rm col}&\approx 1.65\times 10^{-5}\ {\rm Hz}\left(\frac{g_*}{100}\right)^{1/6}\frac{0.62}{1.8-0.1v_w+v_w^2}\left(\frac{T_*}{100\ \rm GeV}\right)\frac{\beta}{H_*}\\
    &\approx 2.05\times 10^9\ {\rm Hz} \ \frac{v_w}{1.8-0.1v_w+v_w^2}(\beta_{\rm PBH,i}')^{1/3}\left(\frac{\rm g}{M_{\rm PBH}}\right)^{1/2}.
\end{align}
The fraction of latent heat converted into the kinetic energy of the bubble walls is \cite{Kamionkowski:1993fg}
\begin{equation}
    \kappa_{\rm col}=\frac{1}{1+0.715\alpha_*}\left(0.715\alpha_*+\frac{4}{27}\sqrt{\frac{3\alpha_*}{2}}\right).
\end{equation}

The GW spectrum from sound wave is
\begin{align}
    \Omega_{\rm sw}&\approx 5.9\times 10^{-6}\left(\frac{100}{g_*}\right)^{1/3}\left(\frac{\beta}{H_*}\right)^{-1}\left(\frac{\kappa_{\rm sw}\alpha_*}{1+\alpha_*}\right)^2v_wS_{\rm sw}(f)\Upsilon\\
    &\approx 3.54\times 10^{-7}\alpha_*^{-1/4}\eta_{\chi}^{-1}\left(\frac{\kappa_{\rm sw}\alpha_*}{1+\alpha_*}\right)^2\left(\frac{g_*}{100}\right)^{-5/12}(\beta_{\rm PBH,i}')^{2/3}S_{\rm sw}(f)\Upsilon,
\end{align}
where the spectral shape is
\begin{equation}
    S_{\rm sw}(f)=\left(\frac{f}{f_{\rm peak,sw}}\right)^3\left(\frac{7}{4+3(f/f_{\rm peak,sw})^2}\right)^{7/2},
\end{equation}
the peak frequency is
\begin{align}
    f_{\rm peak,sw}&\approx 1.9\times 10^{-5}\ \rm Hz\left(\frac{g_*}{100}\right)^{1/6}\left(\frac{T_*}{100\ \rm GeV}\right)\left(\frac{\beta}{H_*}\right)v_w^{-1}\\
    &\approx 3.8\times 10^9\ \rm Hz\ (\beta_{\rm PBH,i}')^{1/3}\left(\frac{\rm g}{M_{\rm PBH}}\right)^{1/2}.\label{eq:f_peak,sw}
\end{align}
The fraction of latent heat converted into the bulk motion of plasma is \cite{Espinosa:2010hh}
\begin{equation}
    \kappa_{\rm sw}=\frac{\sqrt{\alpha_*}}{0.135+\sqrt{0.98+\alpha_*}}.
\end{equation}
The suppression factor due to the finite lifetime of sound wave is given by \cite{Guo:2020grp}
\begin{equation}
    \Upsilon=1-\frac{1}{\sqrt{1+2\tau_{\rm sw}H_*}}=1-\left[1+0.41\kappa_{\rm sw}^{-1/2}\alpha_*^{-3/4}\eta_\chi^{-1}\left(\frac{g_*}{100}\right)^{-1/12}(\beta_{\rm PBH,i}')^{2/3} \right]^{-1/2}.
\end{equation}

The GW spectrum from turbulence is
\begin{align}
    \Omega_{\rm turb}&\approx 7.46\times 10^{-4}\left(\frac{100}{g_*}\right)^{1/3}\left(\frac{\kappa_{\rm turb}\alpha_*}{1+\alpha_*}\right)^{3/2}v_w\left(\frac{\beta}{H_*}\right)^{-1}S_{\rm turb}(f)\\
    &\approx 4.48\times 10^{-5}\alpha_*^{-1/4}\eta_\chi^{-1}\left(\frac{\kappa_{\rm turb}\alpha_*}{1+\alpha_*}\right)^{3/2}\left(\frac{g_*}{100}\right)^{-5/12}(\beta_{\rm PBH,i}')^{2/3}S_{\rm turb}(f),
\end{align}
where the spectral shape is
\begin{equation}
    S_{\rm turb}(f)=\frac{(f/f_{\rm peak,turb})^3}{[1+(f/f_{\rm peak,turb})]^{11/3}(1+8\pi f/h_*)},
\end{equation}
the peak frequency is
\begin{align}
    f_{\rm peak,turb}&\approx 2.7\times 10^{-5}\ {\rm Hz}\left(\frac{g_*}{100}\right)^{1/6}v_w^{-1}\left(\frac{T_*}{100 \ \rm GeV}\right)\frac{\beta}{H_*}\\
    &\approx 5.4\times 10^9\ \rm Hz\ (\beta_{\rm PBH,i}')^{1/3}\left(\frac{\rm g}{M_{\rm PBH}}\right)^{1/2},
\end{align}
the fraction of latent heat converted into plasma turbulence is $\kappa_{\rm turb}\approx 0.1\kappa_{\rm sw}$ \cite{Borah:2024lml,Hindmarsh:2015qta}, and the frequency at the phase transition time is
\begin{align}
    h_*&\approx 1.65\times 10^{-5}\ \rm Hz\left(\frac{g_*}{100}\right)^{1/6}\left(\frac{T_*}{100\ \rm GeV}\right)\\
    &\approx 1.98\times 10^8\ \rm Hz\ \alpha_*^{-1/4}\eta_\chi^{-1}\left(\frac{g_*}{100}\right)^{-1/12}\beta_{\rm PBH,i}'\left(\frac{\rm g}{M_{\rm PBH}}\right)^{1/2}.
\end{align}

\begin{figure}[h!]
    \centering
    \includegraphics[scale=1]{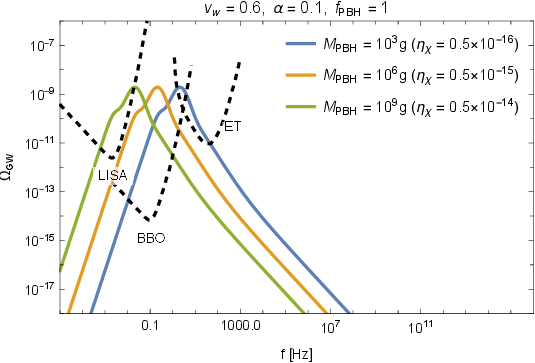}
    \caption{GW spectrum from a FOPT for different PBH masses with the corresponding values of $\eta_\chi$. The fixed parameters are shown on the top. The black dashed contours represent sensitivity regions from different experiments.}
    \label{fig:GW FOPT}
\end{figure}

From the above expressions, we see that the GW signal depends only weakly on $v_w$. Therefore, let us fix this parameter with its typical value: $v_w\sim 0.6$ \cite{Moore:1995ua}. The GW signal is more sensitive to $\alpha_*$ which typically can be of order $O(0.01-1)$. Although $\alpha_*\sim 1$ would give the strongest signal, let us be a bit conservative and pick $\alpha_*\sim 0.1$ as it is typically realized in many models \cite{Caprini:2015zlo}. The peak of GW spectrum also depends on the asymmetry parameter $\eta_\chi$ and the PBH's initial population $\beta_{\rm PBH,i}'$. The latter can be fixed for a chosen mass of PBH and when we demand that PBHs must be entire DM with $f_{\rm PBH}=1$ (see Eq. \ref{eq:beta PBH prime}). Therefore, given the fact that $\Omega_{\rm GW}\propto \eta_\chi^{-1}$, the smallest $\eta_\chi$ would give the strongest GW signal that we hope for. The actual value of $\eta_\chi$ depends on the underlying model and is generally unconstrained \cite{Hong:2020est}\footnote{By definition, $\eta_\chi\lesssim n_\chi^{\rm eq}(T)/s(T)\sim 10^{-3}$ \cite{Kawana:2021tde}, but there is no lower bound. The baryon asymmetry is $\sim 10^{-10}$ but $\chi$ is in the dark sector.}. We find that $\eta_\chi=\{0.5\times 10^{-16},0.5\times 10^{-15},0.5\times 10^{-14}\}$ would give $T_*\approx\{7\times 10^6\ \rm GeV,7\times 10^5\ \rm GeV,7\times 10^4\ \rm GeV\}$ and $\beta/H_*\approx 1$ for $M_{\rm PBH}=\{10^3\ \rm g,10^6\ \rm g,10^9\ \rm g\}$ respectively. Reducing further the value of $\eta_\chi$ would result in too small $\beta/H_*$ that is not viable in most scenarios of FOPT. Therefore, the mentioned set of $\eta_\chi$ is a good benchmark point as it gives the strongest possible GW signal. The total GW spectrum from the FOPT is shown in Fig. \ref{fig:GW FOPT} for different PBH's masses and $f_{\rm PBH}=1$. For an example of individual contributions to the GW spectrum, see Appendix B. 

From Fig. \ref{fig:GW FOPT}, we have some comments:
\begin{itemize}
    \item For the same benchmark values of PBH's final mass and population that we used in the case of inflation, we see that the GW signals now peak at lower frequency ($f_{\rm peak}\sim O(0.01-1)\ \rm Hz$) and have sufficiently large amplitudes ($\Omega_{\rm GW,peak}\sim 10^{-9}$) that can be detected by future experiments.
    \item The peak frequency is inversely proportional to the mass, which is similar to the inflation case. However, unlike inflation, it also depends on the PBH's final population (see Eqs. \ref{eq:f_peak,sw} and \ref{eq:beta PBH prime}). We observe that, although the magnitude of the GW peak depends on parameter choice of the underlying FOPT, the peak frequency is a robust prediction when we look for a specific mass and population of PBH.
\end{itemize}

In the above results, we chose the smallest possible $\eta_\chi$ such that we can have the maximum possible GW signal for each PBH mass. In general, even if it happens that the detectors give some smaller GW peaks, larger values of $\eta_\chi$ can still accommodate this scenario. Furthermore, the GW signal can also give insights into the underlying parameters space of the FOPT that made PBHs. 

\begin{figure}[h!]
    \centering
    \includegraphics[scale=0.82]{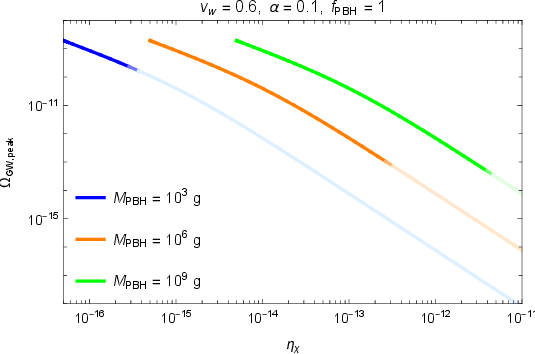}
    \includegraphics[scale=1]{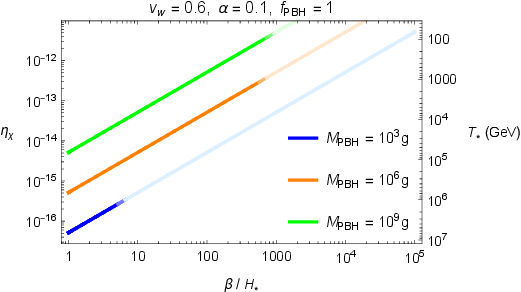}
    \caption{\textbf{\textit{Left panel:}} $\Omega_{\rm GW,peak}$ as a function of $\eta_\chi$ for different PBH masses. Other fixed parameters are shown on the top of the panel. The light portion of the curves cannot be detected by current proposed experiments. \textbf{\textit{Right panel}}: For each value of $\eta_\chi$ determined from the left panel, $\beta/H_*$ and $T_*$ can be inferred. Note the inverse scale of $T_*$.}
    \label{fig:FOPT parameters}
\end{figure}

In  the left panel of Fig. \ref{fig:FOPT parameters}, we show $\Omega_{\rm GW,peak}$ as a function of $\eta_\chi$ for different PBH masses. Other fixed benchmark parameters are the same as before and are shown on the top of the panel. The light portion of the curves cannot be detected by current proposed experiments\footnote{These just include some major proposals such as LISA, BBO, ET. Of course, if it happens that we can have more experiments with greater sensitivity, then more regions of the parameter space can be probed.}. For each value of $\eta_\chi$ determined from the left panel, we can infer the corresponding values of $\beta/H_*$ and $T_*$, as shown in the right panel of Fig. \ref{fig:FOPT parameters}. Therefore, a detection of such GW could be an indication of burdened PBHs as DM scenario and could also offer insights into the underlying details of FOPT.

Some comments are in order regarding the prospect for the required value of $\eta_\chi$. As mentioned earlier, the asymmetry parameter for the dark sector has no lower bound and therefore it can, in principle, take arbitrarily small value. For example, in a model involving right-handed neutrinos decaying into $\chi$ and some scalar $S$ discussed in \cite{Hong:2020est}, $\eta_\chi\sim 10^{-16}$ can be achieved if $M_S\approx 0.9988M_\nu$. Such small value of $\eta_\chi$ is only necessary for $M_{\rm PBH}=10^3\rm g$ in order to have a sufficiently strong GW signal. For other benchmark such as $M_{\rm PBH}=10^{9}\rm g$, $\eta_\chi$ can be as large as $10^{-12}$ and the signal is still strong enough to be detected (see Fig. \ref{fig:FOPT parameters}). Our generic phenomenological approach is therefore a useful guideline for model builders.

For burdened PBHs with mass inside the DM window, they are not evaporating now and their partial evaporation has already been completed before BBN. Therefore, GW is the only way to probe these PBHs. However, there are two situations where we can have additional complementary probes of the burdened PBHs-as-DM scenario:

(1) If PBHs are locked in binary systems, they can merge and form new PBHs. These fresh PBHs are no longer subject to memory burden and therefore can emit some high-energy gamma rays or neutrinos that should be detected by current experiments \cite{Zantedeschi:2024ram}. The evaporation will be suppressed again after the half-decay time and the PBHs become stabilized.

(2) If PBHs collide with neutron stars (NS), they can produce the fast radio bursts (FRBs) \cite{Kainulainen:2021rbg}. FRBs are short pulses in the radio frequency spectrum whose origin is currently unknown. The first detection of such FRBs was reported in \cite{Lorimer:2007qn} based on the data from the Parkes radio telescope. A possible explanation for this signal is that, when a PBH collide with a NS, a FRB could be produced due to the release of the magnetic field energy at the final stage of the swallowing process. The authors of \cite{Kainulainen:2021rbg} found that PBHs with mass below $10^{22}\rm g$ could reproduce the observed FRB signal, within an order of magnitude estimate. Besides the usual asteroid-mass window, this result also allows for the possibility of burdened PBHs as DM. Now, because the peak frequency of GW is inversely proportional to the mass ($f_{\rm peak,sw}\propto M_{\rm PBH}^{-1/3}$ from Eqs. \ref{eq:f_peak,sw} and \ref{eq:beta PBH prime}), the asteroid-mass window would give a peak at much lower frequency outside of the detectors' sensitivity regions. The most promising case is $M_{\rm PBH}=10^{17}\rm g$ that gives $f_{\rm peak,sw}\sim 10^{-5}\ \rm Hz$, which is barely at the edge of LISA. Therefore, if the observed FRB indeed came from PBH-NS collision and the GW is detected in future experiments at the right frequency that we showed here, then we can say with high probability that burdened PBHs made up the DM.

\section{Conclusion}\label{sec:conclusion}
In this paper, we calculated the GW signals associated with the formation of PBHs in the new mass window below $10^{10}\rm g$ where they can be the entire DM thanks to the quantum memory burden effect. We considered the case of PBHs formed from inflationary perturbation and the case of PBHs formed from collapse of Fermi-balls in a FOPT. The scalar-induced GW spectrum from inflationary perturbation peaks at high frequency and demands novel experimental proposals to validate. On the other hand, the GW spectrum from FOPT peaks at lower frequency with sufficiently large amplitude that can be within the reach of near-future experiments such as LISA, BBO, or ET. Detection of such GW signals in upcoming experiments could be an evidence for quantum memory burden effect as well as the PBHs as DM scenario. It can also offer new insights into the parameter space of the underlying physical mechanisms that made the PBHs.

\section*{Appendix A: Redshift of GW spectrum in generic cosmological backgrounds}
In this appendix, we review how the GW spectrum is redshifted due to the expansion of the Universe. For example, see \cite{Allahverdi:2020bys}. Although we assumed a standard RD phase before BBN in the main text, nonstandard cosmologies may arise in many different models and might be of interest to some readers who want to extend our results here. We imagine a single fluid $\phi$ with the equation of state, $w$, dominates the Universe's expansion at early time and then reheat to radiation before BBN. We assume there is no further entropy injection afterwards. Since $\rho_{\rm GW}\propto a^{-4}$ and $\rho_\phi\propto a^{-3(1+w)}$, the fractional energy density of GW today is
\begin{align}
    \Omega_{\rm GW,0}&=\left(\frac{a_*}{a_0}\right)^4\left(\frac{H_*}{H_0}\right)^2\Omega_{\rm GW,*}\\
    &=\left(\frac{a_*}{a_R}\right)^4\left(\frac{a_R}{a_0}\right)^4\left(\frac{H_*}{H_0}\right)^2\Omega_{\rm GW,*}\\
    &=\left(\frac{H_*}{H_R}\right)^{\frac{-8}{3(1+w)}}\left(\frac{a_R}{a_0}\right)^4\left(\frac{H_*}{H_0}\right)^2\Omega_{\rm GW,*}\\
    &=\left(\frac{H_*}{H_R}\right)^{2\frac{3w-1}{3(w+1)}}\left(\frac{a_R}{a_0}\right)^4\left(\frac{H_R}{H_0}\right)^2\Omega_{\rm GW,*}\\
    &=\left(\frac{H_*}{H_R}\right)^{2\frac{3w-1}{3(w+1)}}\left(\frac{g_{\rm s,0}^{1/3}T_0}{g_{\rm s,R}^{1/3}T_R}\right)^4\left(\frac{\pi^2g_{\rho,R}T_R^4}{45M_{\rm pl}^2H_0^2}\right)\Omega_{\rm GW,*}\\
    &\approx 3.15 \times 10^{-4}g_{\rho,R}^{-1/3}\left(\frac{H_*}{H_R}\right)^{2\frac{3w-1}{3(w+1)}}\Omega_{\rm GW,*},\label{eq: Omega GW w}
\end{align}
where the subscript */R/0 denotes the relevant quantities at the emission/reheating/today time, $M_{\rm pl}$ is the reduced Planck mass and we used the fact that $g_{\rho,R}\sim g_{\rm s,R}$.

For example, an early matter dominated phase (EMD) with $w=0$ would result in
\begin{align}
    \Omega_{\rm GW,0}&\approx 3.15\times 10^{-4}g_{\rho ,R}^{-1/3}\left(\frac{H_*}{H_R}\right)^{-2/3}\Omega_{\rm GW,*}\\
    &=3.15\times 10^{-4}g_{\rho,R}^{-1/3}\left(\frac{a_*}{a_R}\right)\Omega_{\rm GW,*},
\end{align}
Thus, an EMD that lasts seven e-folds from the GW emission time would suppress the GW signal by three orders of magnitude. Note that this result is simply a redshift of GW from emission time to today and we are not considering any other effects such as amplification of GW due to perturbation growth during EMD \cite{Kohri:2018awv}. If $w=1/3$, Eq. \ref{eq: Omega GW w} reduces to the standard result in Eq. \ref{eq: Omega GW RD}. There is a factor of 2 different because the reheating time is defined as $H_R\propto \rho_\phi+\rho_{\rm rad}\sim 2\rho_{\rm rad}$.

The frequency is similarly redshifted as
\begin{align}
    f_0&=\left(\frac{a_*}{a_0}\right)f_*\\
    &=\left(\frac{a_*}{a_R}\right)\left(\frac{a_R}{a_0}\right)f_*\\
    &=\left(\frac{H_*}{H_R}\right)^{\frac{-2}{3(1+w)}}\frac{g_{\rm s,0}^{1/3}T_0}{g_{\rm s,R}^{1/3}T_R}f_*\\
    &\approx 3.63\times 10^{-13}g_{\rho,R}^{-1/3}\left(\frac{\rm GeV}{T_R}\right)\left(\frac{H_*}{H_R}\right)^{\frac{-2}{3(1+w)}}f_*.
\end{align}
For an EMD phase that lasts seven e-folds from the emission time and $T_R=1\ \rm GeV$, the frequency is redshifted towards lower frequency by three orders of magnitude. The standard RD case can again be obtained by taking the limit $a_R\rightarrow a_*$ and $T_R\rightarrow T_*$.

\section*{Appendix B: Comparing different contributions to the GW spectrum from the FOPT}
\begin{figure}[h!]
    \centering
    \includegraphics[scale=1]{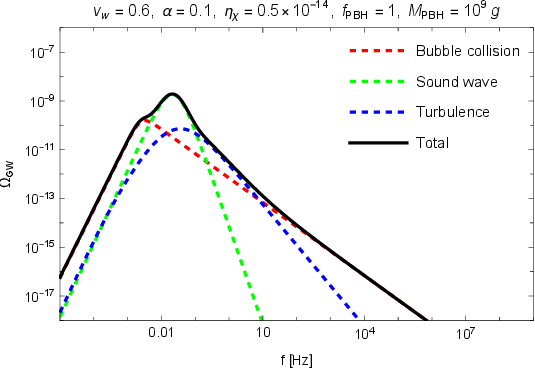}
    \caption{Individual contributions to the total GW spectrum in a FOPT. The chosen benchmark parameters are shown on the top.}
    \label{fig:GW FOPT contributions}
\end{figure}

In this appendix, we compare different contributions to the GW spectrum produced from a FOPT. We consider the benchmark point with PBHs with mass of $10^{9}\rm g$ that can be entire DM $(f_{\rm PBH}=1)$ as an example. The result is shown in Fig. \ref{fig:GW FOPT contributions}. Note that the results in the main text are always the total spectrum.

From Fig. \ref{fig:GW FOPT contributions}, we see that the GW signals from bubble collision, sound wave, and turbulence have three different peak frequencies. The total GW signal is dominated by the sound wave contribution at the peak but is dominated by the bubble collision or turbulence contributions far from the peak.

\bibliographystyle{unsrt}
\bibliography{references}

\begin{thebibliography}{10}

\bibitem{Bertone:2016nfn}
Gianfranco Bertone and Dan Hooper.
\newblock {History of dark matter}.
\newblock {\em Rev. Mod. Phys.}, 90(4):045002, 2018.

\bibitem{Carr:2020gox}
Bernard Carr, Kazunori Kohri, Yuuiti Sendouda, and Jun'ichi Yokoyama.
\newblock {Constraints on primordial black holes}.
\newblock {\em Rept. Prog. Phys.}, 84(11):116902, 2021.

\bibitem{Hawking:1975vcx}
S.~W. Hawking.
\newblock {Particle Creation by Black Holes}.
\newblock {\em Commun. Math. Phys.}, 43:199--220, 1975.
\newblock [Erratum: Commun.Math.Phys. 46, 206 (1976)].

\bibitem{Dvali:2020wft}
Gia Dvali, Lukas Eisemann, Marco Michel, and Sebastian Zell.
\newblock {Black hole metamorphosis and stabilization by memory burden}.
\newblock {\em Phys. Rev. D}, 102(10):103523, 2020.

\bibitem{Dvali:2024hsb}
Gia Dvali, Juan~Sebasti\'an Valbuena-Berm\'udez, and Michael Zantedeschi.
\newblock {Memory burden effect in black holes and solitons: Implications for PBH}.
\newblock {\em Phys. Rev. D}, 110(5):056029, 2024.

\bibitem{Bekenstein:1973ur}
Jacob~D. Bekenstein.
\newblock {Black holes and entropy}.
\newblock {\em Phys. Rev. D}, 7:2333--2346, 1973.

\bibitem{Alexandre:2024nuo}
Ana Alexandre, Gia Dvali, and Emmanouil Koutsangelas.
\newblock {New mass window for primordial black holes as dark matter from the memory burden effect}.
\newblock {\em Phys. Rev. D}, 110(3):036004, 2024.

\bibitem{Thoss:2024hsr}
Valentin Thoss, Andreas Burkert, and Kazunori Kohri.
\newblock {Breakdown of hawking evaporation opens new mass window for primordial black holes as dark matter candidate}.
\newblock {\em Mon. Not. Roy. Astron. Soc.}, 532(1):451--459, 2024.

\bibitem{Haque:2024eyh}
Md~Riajul Haque, Suvashis Maity, Debaprasad Maity, and Yann Mambrini.
\newblock {Quantum effects on the evaporation of PBHs: contributions to dark matter}.
\newblock 4 2024.

\bibitem{Barman:2024iht}
Basabendu Barman, Md~Riajul Haque, and \'Oscar Zapata.
\newblock {Gravitational wave signatures of cogenesis from a burdened PBH}.
\newblock {\em JCAP}, 09:020, 2024.

\bibitem{Papanikolaou:2020qtd}
Theodoros Papanikolaou, Vincent Vennin, and David Langlois.
\newblock {Gravitational waves from a universe filled with primordial black holes}.
\newblock {\em JCAP}, 03:053, 2021.

\bibitem{Papanikolaou:2022chm}
Theodoros Papanikolaou.
\newblock {Gravitational waves induced from primordial black hole fluctuations: the~effect of an extended mass function}.
\newblock {\em JCAP}, 10:089, 2022.

\bibitem{Papanikolaou:2024kjb}
Theodoros Papanikolaou, Xin-Chen He, Xiao-Han Ma, Yi-Fu Cai, Emmanuel~N. Saridakis, and Misao Sasaki.
\newblock {New probe of non-Gaussianities with primordial black hole induced gravitational waves}.
\newblock {\em Phys. Lett. B}, 857:138997, 2024.

\bibitem{He:2024luf}
Xin-Chen He, Yi-Fu Cai, Xiao-Han Ma, Theodoros Papanikolaou, Emmanuel~N. Saridakis, and Misao Sasaki.
\newblock {Gravitational waves from primordial black hole isocurvature: the effect of non-Gaussianities}.
\newblock 9 2024.

\bibitem{Balaji:2024hpu}
Shyam Balaji, Guillem Dom\`enech, Gabriele Franciolini, Alexander Ganz, and Jan Tr\"ankle.
\newblock {Probing modified Hawking evaporation with gravitational waves from the primordial black hole dominated universe}.
\newblock 3 2024.

\bibitem{Bhaumik:2024qzd}
Nilanjandev Bhaumik, Md~Riajul Haque, Rajeev~Kumar Jain, and Marek Lewicki.
\newblock {Memory burden effect mimics reheating signatures on SGWB from ultra-low mass PBH domination}.
\newblock {\em JHEP}, 10:142, 2024.

\bibitem{Carr:1974nx}
Bernard~J. Carr and S.~W. Hawking.
\newblock {Black holes in the early Universe}.
\newblock {\em Mon. Not. Roy. Astron. Soc.}, 168:399--415, 1974.

\bibitem{Kawana:2021tde}
Kiyoharu Kawana and Ke-Pan Xie.
\newblock {Primordial black holes from a cosmic phase transition: The collapse of Fermi-balls}.
\newblock {\em Phys. Lett. B}, 824:136791, 2022.

\bibitem{Ghosh:2022okj}
Diptimoy Ghosh and Arvind~Kumar Mishra.
\newblock {Gravitation wave signal from asteroid mass primordial black hole dark matter}.
\newblock {\em Phys. Rev. D}, 109(4):043537, 2024.

\bibitem{Huang:2022him}
Peisi Huang and Ke-Pan Xie.
\newblock {Primordial black holes from an electroweak phase transition}.
\newblock {\em Phys. Rev. D}, 105(11):115033, 2022.

\bibitem{Xie:2023cwi}
Ke-Pan Xie.
\newblock {Pinning down the primordial black hole formation mechanism with gamma-rays and gravitational waves}.
\newblock {\em JCAP}, 06:008, 2023.

\bibitem{Dvali:2021byy}
Gia Dvali, Florian K\"uhnel, and Michael Zantedeschi.
\newblock {Primordial black holes from confinement}.
\newblock {\em Phys. Rev. D}, 104(12):123507, 2021.

\bibitem{Papanikolaou:2023crz}
Theodoros Papanikolaou.
\newblock {Primordial black holes in loop quantum cosmology: the effect on the threshold}.
\newblock {\em Class. Quant. Grav.}, 40(13):134001, 2023.

\bibitem{Banerjee:2022xft}
Shreya Banerjee, Theodoros Papanikolaou, and Emmanuel~N. Saridakis.
\newblock {Constraining F(R) bouncing cosmologies through primordial black holes}.
\newblock {\em Phys. Rev. D}, 106(12):124012, 2022.

\bibitem{Chianese:2024rsn}
Marco Chianese, Andrea Boccia, Fabio Iocco, Gennaro Miele, and Ninetta Saviano.
\newblock {The light burden of memory: constraining primordial black holes with high-energy neutrinos}.
\newblock 10 2024.

\bibitem{Zantedeschi:2024ram}
Michael Zantedeschi and Luca Visinelli.
\newblock {Memory-Burdened Primordial Black Holes as Astrophysical Particle Accelerators}.
\newblock 10 2024.

\bibitem{Kohri:2024qpd}
Kazunori Kohri, Takahiro Terada, and Tsutomu~T. Yanagida.
\newblock {Induced Gravitational Waves probing Primordial Black Hole Dark Matter with Memory Burden}.
\newblock 9 2024.

\bibitem{Barker:2024mpz}
Will Barker, Benjamin Gladwyn, and Sebastian Zell.
\newblock {Inflationary and Gravitational Wave Signatures of Small Primordial Black Holes as Dark Matter}.
\newblock 10 2024.

\bibitem{Jiang:2024aju}
Yang Jiang, Chen Yuan, Chong-Zhi Li, and Qing-Guo Huang.
\newblock {Constraints on the Primordial Black Hole Abundance through Scalar-Induced Gravitational Waves from Advanced LIGO and Virgo's First Three Observing Runs}.
\newblock 9 2024.

\bibitem{Franciolini:2023osw}
Gabriele Franciolini and Paolo Pani.
\newblock {Stochastic gravitational-wave background at 3G detectors as a smoking gun for microscopic dark matter relics}.
\newblock {\em Phys. Rev. D}, 108(8):083527, 2023.

\bibitem{Planck:2018vyg}
N.~Aghanim et~al.
\newblock {Planck 2018 results. VI. Cosmological parameters}.
\newblock {\em Astron. Astrophys.}, 641:A6, 2020.
\newblock [Erratum: Astron.Astrophys. 652, C4 (2021)].

\bibitem{Carr:1975qj}
Bernard~J. Carr.
\newblock {The Primordial black hole mass spectrum}.
\newblock {\em Astrophys. J.}, 201:1--19, 1975.

\bibitem{Kohri:2018qtx}
Kazunori Kohri and Takahiro Terada.
\newblock {Primordial Black Hole Dark Matter and LIGO/Virgo Merger Rate from Inflation with Running Spectral Indices: Formation in the Matter- and/or Radiation-Dominated Universe}.
\newblock {\em Class. Quant. Grav.}, 35(23):235017, 2018.

\bibitem{Planck:2018jri}
Y.~Akrami et~al.
\newblock {Planck 2018 results. X. Constraints on inflation}.
\newblock {\em Astron. Astrophys.}, 641:A10, 2020.

\bibitem{Cabass:2016ldu}
Giovanni Cabass, Eleonora Di~Valentino, Alessandro Melchiorri, Enrico Pajer, and Joseph Silk.
\newblock {Constraints on the running of the running of the scalar tilt from CMB anisotropies and spectral distortions}.
\newblock {\em Phys. Rev. D}, 94(2):023523, 2016.

\bibitem{Byrnes:2018txb}
Christian~T. Byrnes, Philippa~S. Cole, and Subodh~P. Patil.
\newblock {Steepest growth of the power spectrum and primordial black holes}.
\newblock {\em JCAP}, 06:028, 2019.

\bibitem{Kozaczuk:2021wcl}
Jonathan Kozaczuk, Tongyan Lin, and Ethan Villarama.
\newblock {Signals of primordial black holes at gravitational wave interferometers}.
\newblock {\em Phys. Rev. D}, 105(12):123023, 2022.

\bibitem{Maggiore:2018sht}
Michele Maggiore.
\newblock {\em {Gravitational Waves. Vol. 2: Astrophysics and Cosmology}}.
\newblock Oxford University Press, 3 2018.

\bibitem{Domenech:2021ztg}
Guillem Dom\`enech.
\newblock {Scalar Induced Gravitational Waves Review}.
\newblock {\em Universe}, 7(11):398, 2021.

\bibitem{Yuan:2021qgz}
Chen Yuan and Qing-Guo Huang.
\newblock {A topic review on probing primordial black hole dark matter with scalar induced gravitational waves}.
\newblock {\em iScience}, 24:102860, 2021.

\bibitem{Kohri:2018awv}
Kazunori Kohri and Takahiro Terada.
\newblock {Semianalytic calculation of gravitational wave spectrum nonlinearly induced from primordial curvature perturbations}.
\newblock {\em Phys. Rev. D}, 97(12):123532, 2018.

\bibitem{LISA:2017pwj}
Pau Amaro-Seoane et~al.
\newblock {Laser Interferometer Space Antenna}.
\newblock 2 2017.

\bibitem{Yagi:2011wg}
Kent Yagi and Naoki Seto.
\newblock {Detector configuration of DECIGO/BBO and identification of cosmological neutron-star binaries}.
\newblock {\em Phys. Rev. D}, 83:044011, 2011.
\newblock [Erratum: Phys.Rev.D 95, 109901 (2017)].

\bibitem{Kawamura:2011zz}
Seiji Kawamura et~al.
\newblock {The Japanese space gravitational wave antenna: DECIGO}.
\newblock {\em Class. Quant. Grav.}, 28:094011, 2011.

\bibitem{LIGOScientific:2016wof}
Benjamin~P Abbott et~al.
\newblock {Exploring the Sensitivity of Next Generation Gravitational Wave Detectors}.
\newblock {\em Class. Quant. Grav.}, 34(4):044001, 2017.

\bibitem{Punturo:2010zz}
M.~Punturo et~al.
\newblock {The Einstein Telescope: A third-generation gravitational wave observatory}.
\newblock {\em Class. Quant. Grav.}, 27:194002, 2010.

\bibitem{Moore:2014lga}
C.~J. Moore, R.~H. Cole, and C.~P.~L. Berry.
\newblock {Gravitational-wave sensitivity curves}.
\newblock {\em Class. Quant. Grav.}, 32(1):015014, 2015.

\bibitem{ET:2019dnz}
Michele Maggiore et~al.
\newblock {Science Case for the Einstein Telescope}.
\newblock {\em JCAP}, 03:050, 2020.

\bibitem{Hindmarsh:2015qta}
Mark Hindmarsh, Stephan~J. Huber, Kari Rummukainen, and David~J. Weir.
\newblock {Numerical simulations of acoustically generated gravitational waves at a first order phase transition}.
\newblock {\em Phys. Rev. D}, 92(12):123009, 2015.

\bibitem{Caprini:2015zlo}
Chiara Caprini et~al.
\newblock {Science with the space-based interferometer eLISA. II: Gravitational waves from cosmological phase transitions}.
\newblock {\em JCAP}, 04:001, 2016.

\bibitem{Kamionkowski:1993fg}
Marc Kamionkowski, Arthur Kosowsky, and Michael~S. Turner.
\newblock {Gravitational radiation from first order phase transitions}.
\newblock {\em Phys. Rev. D}, 49:2837--2851, 1994.

\bibitem{Espinosa:2010hh}
Jose~R. Espinosa, Thomas Konstandin, Jose~M. No, and Geraldine Servant.
\newblock {Energy Budget of Cosmological First-order Phase Transitions}.
\newblock {\em JCAP}, 06:028, 2010.

\bibitem{Guo:2020grp}
Huai-Ke Guo, Kuver Sinha, Daniel Vagie, and Graham White.
\newblock {Phase Transitions in an Expanding Universe: Stochastic Gravitational Waves in Standard and Non-Standard Histories}.
\newblock {\em JCAP}, 01:001, 2021.

\bibitem{Borah:2024lml}
Debasish Borah, Suruj Jyoti~Das, and Indrajit Saha.
\newblock {Dark matter from phase transition generated PBH evaporation with gravitational waves signatures}.
\newblock {\em Phys. Rev. D}, 110(3):035014, 2024.

\bibitem{Moore:1995ua}
Guy~D. Moore and Tomislav Prokopec.
\newblock {Bubble wall velocity in a first order electroweak phase transition}.
\newblock {\em Phys. Rev. Lett.}, 75:777--780, 1995.

\bibitem{Hong:2020est}
Jeong-Pyong Hong, Sunghoon Jung, and Ke-Pan Xie.
\newblock {Fermi-ball dark matter from a first-order phase transition}.
\newblock {\em Phys. Rev. D}, 102(7):075028, 2020.

\bibitem{Kainulainen:2021rbg}
Kimmo Kainulainen, Sami Nurmi, Enrico~D. Schiappacasse, and Tsutomu~T. Yanagida.
\newblock {Can primordial black holes as all dark matter explain fast radio bursts?}
\newblock {\em Phys. Rev. D}, 104(12):123033, 2021.

\bibitem{Lorimer:2007qn}
D.~R. Lorimer, M.~Bailes, M.~A. McLaughlin, D.~J. Narkevic, and F.~Crawford.
\newblock {A bright millisecond radio burst of extragalactic origin}.
\newblock {\em Science}, 318:777, 2007.

\bibitem{Allahverdi:2020bys}
Rouzbeh Allahverdi et~al.
\newblock {The First Three Seconds: a Review of Possible Expansion Histories of the Early Universe}.
\newblock 6 2020.

\end{thebibliography}

\end{document}